\documentclass[12pt]{iopart}

\usepackage{iopams}  
\usepackage{graphicx}
\usepackage{dcolumn}
\usepackage{bm}
\usepackage{multirow}
\usepackage{braket}
\usepackage{color}
\usepackage{ulem}

\begin{document}

\title{Temperature-driven transition from skyrmion to bubble crystals in centrosymmetric itinerant magnets}

\author{Satoru Hayami}

\address{Department of Applied Physics, The University of Tokyo, Tokyo 113-8656, Japan}
\ead{hayami@ap.t.u-tokyo.ac.jp}
\vspace{10pt}

\begin{abstract}
Interplay between itinerant electrons and localized spins in itinerant magnets gives rise to a variety of noncoplanar multiple-$Q$ spin textures, such as the skyrmion, hedgehog, meron, and vortex. 
We elucidate that another type of multiple-$Q$ state consisting of collinear sinusoidal waves, a magnetic bubble crystal, appears at finite temperatures in a centrosymmetric itinerant electron system. 
The results are obtained for the classical Kondo lattice model with easy-axis single-ion anisotropy on a triangular lattice by a large-scale numerical simulation. 
We find that a finite-temperature topological phase transition between the skyrmion crystal and the bubble crystal occurs by changing the temperature. 
We obtain the minimal key ingredients for inducing the finite-temperature transition by analyzing an effective spin model where it is shown that the synergy between the multiple-spin interaction and magnetic anisotropy plays a significant role. 
\end{abstract}

%
\vspace{2pc}

\submitto{\NJP}
%
%

\section{Introduction}

A magnetic skyrmion, which is characterized by a swirling spin texture to have a nontrivial topological (skyrmion) number, has been extensively studied in condensed matter physics, as it exhibits a number of intriguing physical phenomena that arises from emergent magnetic fields, such as the topological Hall effect~\cite{skyrme1962unified,Bogdanov89,Bogdanov94,nagaosa2013topological}. 
Since the discovery of the periodic array of the magnetic skyrmion dubbed a skyrmion crystal (SkX) in MnSi in 2009~\cite{Muhlbauer_2009skyrmion,Neubauer_PhysRevLett.102.186602}, various types of the SkXs have been observed in metals~\cite{Muhlbauer_2009skyrmion,yu2010real,yu2011near,heinze2011spontaneous,nayak2017discovery}, semiconductors~\cite{yu2011near,Kanazawa_PhysRevLett.106.156603}, and insulators~\cite{seki2012observation,Adams2012,kezsmarki_neel-type_2015,Fujima_PhysRevB.95.180410,bordacs2017equilibrium} in noncentrosymmetric magnets~\cite{Tokura_doi:10.1021/acs.chemrev.0c00297}.
Furthermore, a fertile playground for the SkXs has been established in recent years by detecting them in centrosymmetric magnets~\cite{Saha_PhysRevB.60.12162,kurumaji2019skyrmion,hirschberger2019skyrmion,khanh2020nanometric,Yasui2020imaging}. 
As the topological properties in the swirling spin texture are robust against external stimuli, the SkXs are promising for high-efficient storage and memory applications~\cite{fert2013skyrmions,romming2013writing,finocchio2016magnetic,fert2017magnetic,everschor2018perspective,zhang2020skyrmion}. 

Theoretically, the stabilization mechanisms of the SkXs have also been clarified in various electron and/or spin systems. 
The most intuitive model to stabilize the SkXs is the localized spin model consisting of the ferromagnetic exchange interaction and the Dzyaloshinskii-Moriya (DM) interaction, the latter of which arises from the relativistic spin-orbit coupling without the inversion symmetry~\cite{dzyaloshinsky1958thermodynamic,moriya1960anisotropic}. 
The model at a zero field exhibits a magnetic spiral state originating from the competition between the ferromagnetic and DM interactions where the ferromagnetic (DM) interaction tends to align (tilt) the neighboring spins.  
The system undergoes a magnetic field-induced phase transition to the SkXs, which are formed by superposing three spiral states~\cite{rossler2006spontaneous,Yi_PhysRevB.80.054416}. 
Subsequently, it is shown that the SkXs are also stabilized by the other interactions and fluctuations even without the Dzyaloshinskii-Moriya interaction~\cite{batista2016frustration}.  
For example, the spin model with the frustrated exchange interactions and/or magnetic anisotropy gives rise to the SkXs~\cite{Okubo_PhysRevLett.108.017206,leonov2015multiply,Lin_PhysRevB.93.064430,Hayami_PhysRevB.93.184413,amoroso2020spontaneous}. 
Another example is found in the itinerant electron model which implicitly includes effective multiple-spin interactions arising from the Fermi surface instability~\cite{Martin_PhysRevLett.101.156402,Akagi_PhysRevLett.108.096401,Hayami_PhysRevB.90.060402,Hayami_PhysRevB.95.224424,Ozawa_PhysRevLett.118.147205}. 
Notably, these mechanisms enable the formation of the SkXs with short-period magnetic modulations in contrast to that induced by the DM mechanism where the magnetic modulation usually has a long period owing to the small DM interaction compared to the exchange interaction. 
The small-period SkX might be important for next-generation spintronic applications, as it realizes an energy-efficient information storage based on high density skyrmions~\cite{zhang2020skyrmion}.

Besides the merit of the applications, the frustrated and/or multiple spin interactions can induce a plethora of exotic spin orderings characterized by the superposition of the spin density waves (multiple-$Q$ states)~\cite{hayami2021topological}, e.g., magnetic bubble crystals~\cite{lin1973bubble,Garel_PhysRevB.26.325,takao1983study,Hayami_PhysRevB.93.184413,seo2021spin,Hayami_PhysRevB.103.224418}, vortex crystals~\cite{Momoi_PhysRevLett.79.2081,Kamiya_PhysRevX.4.011023,Wang_PhysRevLett.115.107201,Marmorini2014,Hayami_PhysRevB.94.174420,hayami2020phase,yambe2021skyrmion}, meron crystals~\cite{Hayami_PhysRevLett.121.137202,Bera_PhysRevResearch.1.033109,Hayami_PhysRevB.103.024439,Wang_PhysRevB.103.104408,Utesov_PhysRevB.103.064414,Hayami_PhysRevB.103.054422,hayami2021meron}, hedgehog lattices~\cite{Binz_PhysRevB.74.214408,Park_PhysRevB.83.184406,Okumura_PhysRevB.101.144416,grytsiuk2020topological,Shimizu_PhysRevB.103.054427,Aoyama_PhysRevB.103.014406}, and chirality density wave~\cite{Solenov_PhysRevLett.108.096403,Ozawa_doi:10.7566/JPSJ.85.103703,Hayami_PhysRevB.94.024424,Shimokawa_PhysRevB.100.224404,yambe2020double}. 
Indeed, the emergence of these peculiar magnetic states including the SkXs has been suggested/observed in various magnetic materials, such as the vortex crystals in MnSc$_2$S$_4$~\cite{Gao2016Spiral,gao2020fractional}, CeAuSb$_2$~\cite{Marcus_PhysRevLett.120.097201,Seo_PhysRevX.10.011035,seo2021spin}, and Y$_3$Co$_8$Sn$_4$~\cite{takagi2018multiple}, the SkXs in EuPtSi~\cite{kakihana2018giant,kaneko2019unique,tabata2019magnetic,kakihana2019unique,hayami2021field}, Gd$_2$PdSi$_3$~\cite{Saha_PhysRevB.60.12162,kurumaji2019skyrmion,sampathkumaran2019report,Hirschberger_PhysRevLett.125.076602,Hirschberger_PhysRevB.101.220401,Kumar_PhysRevB.101.144440,spachmann2021magnetoelastic}, Gd$_3$Ru$_4$Al$_{12}$~\cite{chandragiri2016magnetic,Nakamura_PhysRevB.98.054410,hirschberger2019skyrmion,Hirschberger_10.1088/1367-2630/abdef9}, and GdRu$_2$Si$_2$~\cite{khanh2020nanometric,Yasui2020imaging}, and the hedgehog lattices in MnSi$_{1-x}$Ge$_{x}$~\cite{tanigaki2015real,kanazawa2017noncentrosymmetric,fujishiro2019topological,Kanazawa_PhysRevLett.125.137202} and SrFeO$_3$~\cite{Ishiwata_PhysRevB.84.054427,Ishiwata_PhysRevB.101.134406,Rogge_PhysRevMaterials.3.084404,Onose_PhysRevMaterials.4.114420}. 
Thus, further intriguing topological magnetic states and their related phase transitions are expected in these systems, which opens up a possibility of realizing novel functional materials.   
Meanwhile, as the investigation of such magnetic states requires a tremendous computational cost, especially for the electron systems, the effect of thermal fluctuations on the SkXs and other topological spin states is not fully understood~\cite{hayami2020phase}. 

In the present study, we investigate the stability of the SkX in centrosymmetric itinerant magnets at finite temperatures. 
By considering the classical Kondo lattice model with an easy-axis single-ion anisotropy on a triangular lattice and performing large-scale Langevin dynamics simulations combined with the kernel polynomial method (KPM-LD)~\cite{Barros_PhysRevB.88.235101}, we find that the system exhibits a finite-temperature phase transition between the SkX described by the triple-$Q$ spiral waves and the bubble crystal described by the triple-$Q$ collinear sinusoidal waves. 
This phase transition is regarded as the topological phase transition where the scalar chirality is turned on and off. 
We show that such a finite-temperature phase transition between the SkX and the bubble crystal is captured by an effective spin model derived from the Kondo lattice model, suggseting that the interplay between the multiple-spin interactions and magnetic anisotropy is essential to the transition. 
Our results indicate that temperature fluctuations can be a source of multiple-$Q$ states and drive a further topological phase transition in the skyrmion-hosting itinerant magnets.

\section{Results and discussion}

We discuss the stability of the SkX in itinerant magnets by performing large-scale numerical simulations. 
We show the result in the Kondo lattice model on a triangular lattice in section~\ref{sec:Kondo lattice model} and in the effective spin model in section~\ref{sec:Effective spin model}. 

\subsection{Classical Kondo lattice model}
\label{sec:Kondo lattice model}

To analyze the finite-temperature effect on the SkX in centrosymmetric itinerant magnets, we consider the classical Kondo lattice model (or the double exchange model) on the triangular lattice, where the previous studies have shown that the SkX is stabilized at zero temperature~\cite{Ozawa_PhysRevLett.118.147205,Hayami_PhysRevB.99.094420,hayami2021locking}. 
The Kondo lattice model consists of itinerant electrons and localized spins, whose Hamiltonian is given by 
\begin{eqnarray}
\label{eq:Ham}
\mathcal{H}= -\sum_{i,j, \sigma}t_{ij}c^{\dagger}_{i\sigma}c_{j\sigma}+ J \sum_i \bi{s}_i \cdot \bi{S}_i - A \sum_i (S_i^z)^2 -H \sum_i S_i^z, 
\end{eqnarray}
where $c^{\dagger}_{i\sigma}$ ($c_{i\sigma}$) is a creation (annihilation) operator of an itinerant electron at site $i$ and spin $\sigma$. 
The first term in (\ref{eq:Ham}) stands for the kinetic energy of itinerant electrons where $t_{ij}$ is a hopping parameter between sites $i$ and $j$. 
The second term in (\ref{eq:Ham}) stands for the exchange coupling between itinerant electron spins $\bi{s}_i = (1/2)\sum_{\sigma, \sigma'}c^{\dagger}_{i\sigma}\bm{\sigma}_{\sigma\sigma'}c_{i \sigma'}$ and localized spins $\bi{S}_i$, where $J$ is a coupling constant and $\bm{\sigma}$ is the vector of Pauli matrices. 
We consider the localized spin $\bi{S}_i$ as the classical one with $|\bi{S}_i|=1$ for simplicity. 
The third term in (\ref{eq:Ham}) represents the easy-axis single-ion anisotropy $A>0$. 
The fourth term in (\ref{eq:Ham}) represents the Zeeman coupling to localized spins under an external magnetic field along the $z$ direction. 

The model in (\ref{eq:Ham}) shows a plethora of multiple-$Q$ states including the SkXs by changing $t_{ij}$, $J$, $A$, $H$, and the chemical potential $\mu$~\cite{Martin_PhysRevLett.101.156402,Akagi_JPSJ.79.083711,Ozawa_PhysRevLett.118.147205,Hayami_PhysRevB.99.094420}. 
Among them, we focus on a parameter set where two SkXs are robustly stabilized in the ground state by choosing the nearest-neighbor hopping $t_1=1$, the third nearest-neighbor hopping $t_3=-0.85$, $J=1$, and $\mu=-3.5$; the SkX with the skyrmion number of two (one) appears at the zero (finite) magnetic field for $A=0$~\cite{Ozawa_PhysRevLett.118.147205,hayami2021locking}. 
The emergent SkXs are characterized by the superposition of the triple-$Q$ spin density waves with $\bi{Q}_1=(\pi/3,0)$, $\bi{Q}_2=(-\pi/6,\sqrt{3}\pi/6)$, and $\bi{Q}_3=(-\pi/6,-\sqrt{3}\pi/6)$, which are dictated by the peaks of the bare susceptibility of itinerant electrons. 
By introducing $A$, the stability region of the SkX with the skyrmion number of one under the magnetic fiend is considerably extended compared to that with the number of two~\cite{Hayami_PhysRevB.99.094420}. 
We here examine the temperature effect on the SkX with the skyrmion number of one in the presence of $A$, since all the observed SkXs in itinerant magnets correspond to the SkX with the number of one~\cite{kurumaji2019skyrmion,hirschberger2019skyrmion,khanh2020nanometric}. 
The temperature effect on the SkXs for $A=0$ has been investigated by the author and his collaborators where it was shown that the SkX changes into the topologically-trivial triple-$Q$ state breaking the threefold rotational symmetry at finite temperatures~\cite{hayami2020phase}.

The behavior of the finite temperature is investigated by performing the KPM-LD simulation, which is one of the efficient methods in the itinerant electron model~\cite{Weis_RevModPhys.78.275,Barros_PhysRevB.88.235101}. 
This method is used for the large-scale system where the partition function consists of itinerant electrons coupled to classical boson fields, which has been applied to similar itinerant electron models~\cite{Barros_PhysRevB.88.235101,Barros_PhysRevB.90.245119,Ozawa_doi:10.7566/JPSJ.85.103703,Ozawa_PhysRevLett.118.147205,Wang_PhysRevLett.117.206601,Ozawa_PhysRevB.96.094417,Chern_PhysRevB.97.035120,ono2019photoinduced}. 
In the present Kondo lattice Hamiltonian with the classical localized spins in (\ref{eq:Ham}), the partition function is described by $Z={\rm Tr}_{\{\bm{S}_i\}}{\rm Tr}_{c}\exp [-(\mathcal{H}-\mu N)/T]$ (we set the Boltzmann constant as unity), where $\{\bm{S}_i\}$ is the localized spin configuration and $N$ is the total number operator for the itinerant electrons.
Taking the trace over the itinerant electron degree of freedom, the partition function is given by $Z={\rm Tr}_{\{\bm{S}_i\}}\exp [-F_{c}(\{\bm{S}_i\})/T]$, where $F_{c}(\{\bm{S}_i\})$ is the free energy for a given localized spin configuration and is given by $F_{c}(\{\bm{S}_i\})=-T\int {\rm d}\omega \rho(\omega) \ln (1+e^{-(\omega-\mu)/T})$. 
Here, $\rho(\omega)$ is the density of states of $\mathcal{H}$ under $\{\bm{S}_i\}$, which is evaluated by the kernel
polynomial method~\cite{Weis_RevModPhys.78.275}. 
We expand the density of states by up to 2000th order of Chebyshev polynomials with $16^2$ random vectors~\cite{Tang12}. 
Meanwhile, ${\rm Tr}_{\{\bm{S}_i\}}$ is evaluated by the spin configuration generated by the Langevin equation, where we adopt a projected Heun scheme~\cite{Mentink10} for 1000-5000 steps with the time interval $\Delta \tau =2$. 
We start the simulation from initial states with random spin configurations for the system sizes with $N=L^2$ at $L=96$ and $120$ with periodic boundary conditions in both directions. 
The results at different temperatures are obtained independently starting from different random spin configurations.

\begin{figure}[t!]
\begin{center}
\includegraphics[width=0.5\hsize]{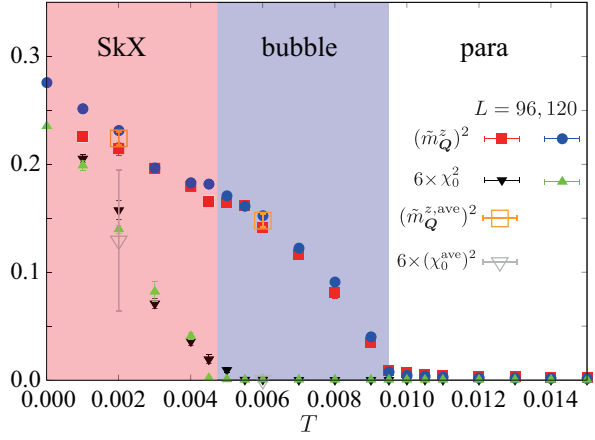} 
\caption{
\label{fig:Tdep}
Temperature ($T$) dependence of the averaged magnetic moments over the vectors $\bm{Q}=\{\bm{Q}_1, \bm{Q}_2, \bm{Q}_3, \cdots \}$ (see also the text), $\tilde{m}^z_{\bm{Q}}$, and the scalar chirality, $\chi_0$, obtained from the KPM-LD simulations in the Kondo lattice model on the triangular lattice for $L=96$ and $120$. 
The averaged values for different seventeen random seeds, $\tilde{m}^{z, {\rm ave}}_{\bm{Q}}$ and $\chi^{\rm ave}_0$, are plotted at $T=0.002$ and $T=0.006$. 
``para" represents the paramagnetic state. 
}
\end{center}
\end{figure}

\begin{figure}[t!]
\begin{center}
\includegraphics[width=0.9\hsize]{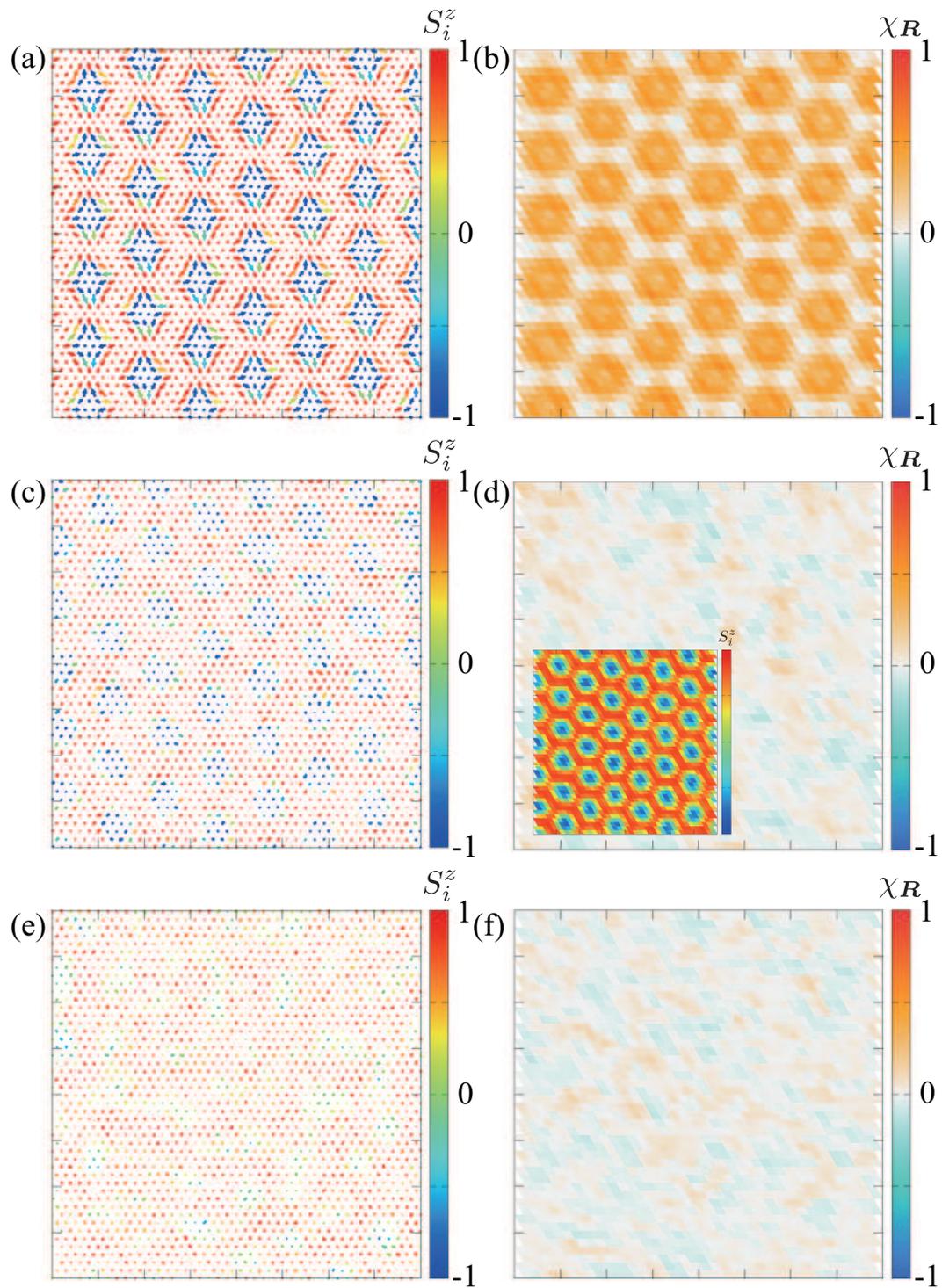} 
\caption{
\label{fig:spin}
Snapshots of the real-space spin [(a), (c), and (e)] and chirality [(b), (d), and (f)] configurations in the (a), (b) SkX at $T=0.002$, (c), (d) bubble crystal at $T=0.006$, and (e), (f) paramagnetic state at $T=0.01$. 
In (a), (c), and (e), the arrows represent the magnetic moments and the color represents the $z$-spin component. 
In the inset of (d), the contour plot of the $z$-spin component in (c) is shown. 
}
\end{center}
\end{figure}

Figure~\ref{fig:Tdep} shows the result at $A=0.007$ and $H=0.008$ obtained by the KPM-LD simulations for the system size with $L=96$ and $120$. 
To identify the magnetic phases, we compute the magnetic moments with wave vector $\bm{q}$, 
\begin{eqnarray}
m^{\alpha}_{\bm{q}}&=\sqrt{\frac{S^{\alpha}(\bm{q})}{N}}, \\
S^{\alpha}(\bm{q})&=\frac{1}{N}\sum_{ij} S^\alpha_i S^\alpha_j e^{i \bm{q}\cdot (\bm{r}_i-\bm{r}_j)}
\end{eqnarray}
where $S^{\alpha}(\bm{q})$ is the $\alpha$ component of the spin structure factor. 
We also calculate the spin scalar chirality as
\begin{eqnarray}
\chi_{\bm{q}}=\sqrt{\frac{S_{\chi}(\bm{q})}{N}}, \\
S_{\chi}(\bm{q})= \frac{1}{N}\sum_{\mu}\sum_{\bm{R},\bm{R}' \in \mu}  \chi_{\bm{R}}
\chi_{\bm{R}'} e^{i \bm{q}\cdot (\bm{R}-\bm{R}')}, 
\end{eqnarray}
where $\mu=(u, d)$ represent upward and downward triangles, respectively, and $\bm{R}$ and $\bm{R}'$ represent the position vectors at the centers of triangles; 
$\chi_{\bm{R}}= \bm{S}_j \cdot (\bm{S}_k \times \bm{S}_l)$ is the local spin chirality at $\bm{R}$, where $j,k,l$ are the sites located at the triangle vertices at $\bm{R}$ in the anticlockwise order. 
The nonzero uniform component of $\chi_{\bm{q}}$, i.e., $\chi_{0}$, can indicate the magnetic phase with the topologically nontrivial spin texture, such as the SkX. 
It is noted that the long-range order in terms of the $xy$ spin is limited to zero temperature in the present two-dimensional system owing to the Mermin-Wagner theorem~\cite{Mermin_PhysRevLett.17.1133}, while the $z$ spin and scalar chirality can order at finite temperatures. 

As shown in figure~\ref{fig:Tdep}, the system has nonzero $\chi_{0}$ at zero temperature. 
While increasing temperature, $\chi_{0}$ gradually decreases and vanishes around $T_1 \sim 0.0045$, which indicates the topological phase transition regarding the spin scalar chirality degree of freedom. 
In other words, this phase transition means the spontaneous $Z_2$ mirror symmetry breaking of the Kondo lattice Hamiltonian according to the ordering of the scalar chirality.
Meanwhile, the spin moment $\tilde{m}^z_{\bm{Q}}$ takes a finite value above $T_1$ where $\tilde{m}^z_{\bm{Q}}$ is defined as 
\begin{equation}
\label{eq:mz}
\tilde{m}^z_{\bm{Q}} = \sqrt{\frac{1}{N}\sum_{\nu,\eta,n} S^z(\bi{Q}_{\nu}-\bi{q}^n_{\nu\eta}) }, 
\end{equation}
where $\bi{Q}_1$-$\bm{Q}_3$ represents the ordering vectors that the bare susceptibility of itinerant electrons is maximized. 
Although the magnetic state is characterized by the state with the multiple peaks at $\bi{Q}_\nu$ in the case of $A=0$~\cite{Ozawa_PhysRevLett.118.147205}, the peak positions can be slightly moved from $\bi{Q}_\nu$ when considering the effect of $A$ and $T$. 
Indeed, such situations have been found in the frustrated magnets with the strong easy-axis anisotropy~\cite{Hayami_PhysRevB.93.184413} and the axial next-nearest-neighbor Ising model where the anisotropy is infinitesimally large~\cite{Elliott_PhysRev.124.346,Fisher_PhysRevLett.44.1502,selke1988annni}. 
To take into account such a change of the peak positions, we introduce $\bm{q}^{n}_{\nu\eta}$ in (\ref{eq:mz}) for the $n$th-neighbor displacement vectors measured from $\bi{Q}_\nu$ ($\eta$ is the number of the $n$th-neighbor displacement vectors), e.g., $\eta=0$ for $n=0$ and $\eta=1$-$6$ for $n=1$-$3$. 
We here take the summation up to $n=3$. 
The transition temperature where $\tilde{m}^z_{\bm{Q}}$ vanishes is around $T_2 \sim 0.009$, which is clearly higher than $T_1$. 
This indicates that the phase transition at $T_2$ means the ordering of the $z$ spin component caused by the single-ion anisotropy.
Thus, the transitions in terms of the spin and chirality occur at different temperatures in the model in (\ref{eq:Ham}). 
Such a separation between spin and chirality degrees of freedom has been discussed in various itinerant electron models, such as the Hubbard model on the kagome lattice~\cite{Udagawa_PhysRevLett.104.106409,udagawa2010chirality} and the Kondo lattice model without the magnetic anisotropy on the triangular lattice~\cite{Kato_PhysRevLett.105.266405,hayami2020phase}. 
The reason why $T_1$ and $T_2$ are different might be attributed to the presence of two different energy scales arising from the effective magnetic interaction mediated by itinerant electrons and the single-ion anisotropy.
Indeed, similar phase transitions are found in the effective spin model derived to include these two factors, as discussed in section~\ref{sec:Effective spin model}.

To examine the spin textures in each phase, we plot the real-space spin and chirality configurations obtained by the KPM-LD simulations in figure~\ref{fig:spin}. 
To integrate out the short-wavelength fluctuations, we average over 200-400 spin configurations for different time steps. 
In the low-temperature phase at $T=0.002$, the SkX spin texture is found in figure~\ref{fig:spin}(a); the skyrmion cores at $S^z_i \sim -1$ form a triangular lattice with the lattice constant $4\pi/(\sqrt{3}|\bi{Q}_{\nu}|)$, which indicates the triple-$Q$ peak structures at $\bi{q} \simeq \bi{Q}_1$, $\bi{Q}_2$, and $\bi{Q}_3$ with equal intensities. 
This phase also accompanies a uniform scalar chirality, as shown in figure~\ref{fig:spin}(b), where the skyrmion number becomes $\pm 1$. 
This SkX has the degeneracy with respect to the helicity and vorticity, which means that this state can take both positive and negative chirality depending on the initial spin configuration, which is similar to that in frustrated magnets~\cite{Okubo_PhysRevLett.108.017206,leonov2015multiply,Lin_PhysRevB.93.064430,Hayami_PhysRevB.93.184413}. 
It is noted that the degeneracy between the SkXs with positive and negative scalar chirality is lifted if the system has the Dzyaloshinskii-Moriya interaction~\cite{bera2021skyrmions} and bond-dependent anisotropic interaction~\cite{amoroso2020spontaneous,Hayami_doi:10.7566/JPSJ.89.103702,Hirschberger_10.1088/1367-2630/abdef9,Hayami_PhysRevB.103.054422,yambe2021skyrmion,amoroso2021tuning} that might not be neglected in real materials. 

Figures~\ref{fig:spin}(c) and \ref{fig:spin}(d) represent the spin and chirality distributions in the intermediate temperature region, respectively, where $\tilde{m}^z_{\bm{Q}}$ is nonzero but $\chi_0$ vanishes. 
The spin configuration in figure~\ref{fig:spin}(c) shows the almost collinear-type spin texture, which results in no scalar chirality in figure~\ref{fig:spin}(d). 
Nevertheless, the $z$ component of the spin moment still forms the triangular lattice, which is clearly found in the inset of figure~\ref{fig:spin}(d). 
Thus, the triple-$Q$ collinear spin texture is realized in the intermediate region, which corresponds to the magnetic bubble crystal. 
The bubble core disappears in the paramagnetic state above $T_2$, as shown in figures~\ref{fig:spin}(e) and \ref{fig:spin}(f).

The above result indicates that the finite-temperature phase transition between the SkX and the bubble crystal is characterized by nonzero $\chi_0$. 
Both the states are characterized by the triple-$Q$ spin density waves, although the type of the individual spin density waves is different with each other. 
On the one hand, the SkX consists of the three spirals whose plane is lied along the $xz$ or $yz$ plane, which is given by 
\begin{eqnarray}
\bi{S}^{x,y}_i &\simeq  & {I_{xy}} \sum_{\nu=1-3} \sin{(\bi{Q}_{\nu} \cdot \bi{r}_i+ \theta_{\nu})} \bi{e}_{\nu},
\nonumber \\
S^z_i &\simeq& m_z - {I_z} \sum_{\nu=1-3} \cos{(\bi{Q}_{\nu} \cdot \bi{r}_i+ \theta_{\nu})} \bi{e}_{\nu}, 
\label{sky}
\end{eqnarray}
where $\bi{e}_1={\hat \bi{x}}$, $\bi{e}_2=-{\hat \bi{x}}/2 + \sqrt{3}{\hat \bi{y}}/2$, and $\bi{e}_3=-{\hat \bi{x}}/2 - \sqrt{3}{\hat \bi{y}}/2$ (${\hat \bi{x}}$ and ${\hat \bi{y}}$ are the unit vectors along the $x$ and $y$ directions). $I_{xy}$, $I_z$, $m_z$, and $\theta_\nu$ depends on the model parameters. 
On the other hand, the bubble crystal is described by the three sinusoidal waves which oscillates along the $z$ direction as 
\begin{eqnarray}
\bi{S}^{x,y}_i & \simeq & 0,
\nonumber \\
S^z_i &\simeq& m_z - {I_z} \sum_{\nu=1-3} \cos{(\bi{Q}_{\nu} \cdot \bi{r}_i+ \theta_{\nu})} \bi{e}_{\nu}. 
\end{eqnarray}
Thus, one can find that the SkX spin texture reduces to the bubble one in the limit of $I_{xy} \to 0$, which indicates that the scalar chirality is largely suppressed in the bubble crystal. 

Finally, let us comment on numerical simulations. 
Although we have adopted the efficient KPM-LD simulations, which roughly reduce the computational cost of the conventional Monte Carlo sampling with the exact diagonalization by $N^3$, it still needs a huge computational cost for the large system sizes with $N \sim 10^4$. 
Depending on the initial spin configurations, the spin states are easily trapped by the multi-domain structures. 
In order to show the variance for the different initial spin configurations at the same temperature, we show the averaged values of $\tilde{m}^z_{\bm{Q}}$ and $\chi_0$ denoted as $\tilde{m}^{z, {\rm ave}}_{\bm{Q}}$ and $\chi^{\rm ave}_0$, respectively, by performing seventeen independent simulations in the SkX phase at $T=0.002$ and in the bubble crystal phase at $T=0.006$, as shown in figure~\ref{fig:Tdep}. 
The result at $T=0.006$ shows a good convergent behavior, while that at $T=0.002$ shows a large variance, especially for $\chi^{\rm ave}_0$. 
This is attributed to the spin and/or chiral domain structures, the latter of which arises from the degeneracy between the SkX with the negative and positive chiralities.
Thus, although a clear signal of the transition between the SkX and bubble crystal is found in the present study, a more careful finite-size scaling is required to examine the nature of the phase transition, which will be left for future study.

\subsection{Effective spin model}
\label{sec:Effective spin model}

To investigate key ingredients for the finite-temperature transition between the SkX and bubble crystal, we investigate an effective spin model in the Kondo lattice model. 
The effective spin model is derived from the perturbation expansion in terms of $J$~\cite{Hayami_PhysRevB.95.224424}, which is given by 
\begin{eqnarray}
\label{eq:HamJK}
\mathcal{H}=& 2\sum_\nu
\left[ -\tilde{J}\bm{S}_{\bm{Q}_{\nu}} \cdot \bm{S}_{-\bm{Q}_{\nu}}
+\frac{K}{N} (\bm{S}_{\bm{Q}_{\nu}} \cdot \bm{S}_{-\bm{Q}_{\nu}})^2 \right] \nonumber \\
&-A \sum_{i} (S^z_i)^2-H \sum_i S_i^z
,   
\end{eqnarray}
where the term in the square bracket corresponds to the effective interaction arising from the spin-charge coupling in the Kondo lattice model, while the second and third terms are the same as the third and fourth terms in (\ref{eq:Ham}). 
The first term represents the spin interactions in momentum space; $\tilde{J}$ is the second-order contribution with respect to $J$, while $K$ is the fourth-order contribution with respect to $J$, both of which are positive. 
We consider the specific $\bm{q}$ components of the spin interactions in momentum space where $\bm{S}_{\bi{q}} = (1/\sqrt{N})\sum_{i}\bm{S}_i e^{-i \bm{q} \cdot \bm{r}_i}$ is the Fourier component of $\bm{S}_i$; we only take into account the interactions at $\bm{Q}_1$, $\bm{Q}_2$, and $\bm{Q}_3$, which are dictated by the peaks of the bare susceptibility of itinerant electrons in the original Kondo lattice model. 
This model is contrast to a spin model with similar biquadratic interactions in real space, which also describes the multiple-$Q$ states~\cite{momoi1997possible,heinze2011spontaneous,Simon_PhysRevMaterials.4.084408,bera2021skyrmions}. 
We set $\tilde{J}=1$ as the energy unit of the model in (\ref{eq:HamJK}). 
The other parameters are chosen at $K=0.5$, $A=0.3$, and $H=0.7$ so as to stabilize the SkX with the skyrmion number of one at zero temperature, and the system size is taken at $L=96$. 

The following result is calculated by performing simulated annealing from high temperature, which is based on the standard Metropolis local updates in real space. 
Similar to the previous studies~\cite{Hayami_PhysRevB.95.224424,hayami2020multiple,Hayami_PhysRevB.103.024439}, we gradually reduce the temperature with the rate $\alpha=0.99995$-$0.99999$ step by step, starting from the initial temperature $T_0=1.5$ for a random spin configuration. 
At the target temperature, we perform $10^5$-$10^6$ Monte Carlo sweeps for measurements after equilibration. 
The following data at different temperature are obtained for the different initial spin configurations. 
The calculation of the effective spin model can be performed by a considerably smaller computational cost than that by the KPM-LD simulation of the Kondo lattice model.

\begin{figure}[t!]
\begin{center}
\includegraphics[width=0.5\hsize]{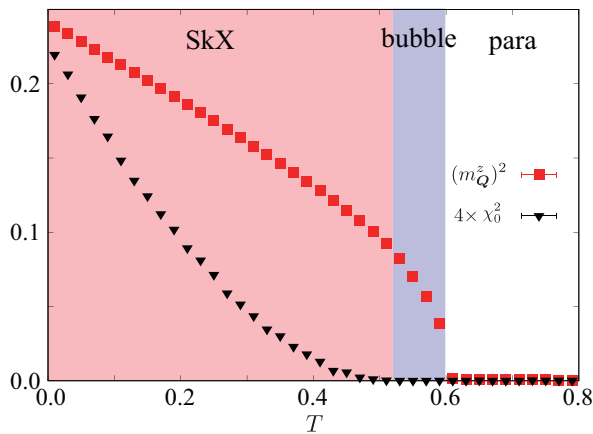} 
\caption{
\label{fig:Tdep_JK}
$T$ dependence of the magnetic moments with the averaged ordering vectors over $\bm{Q}=\{\bm{Q}_1, \bm{Q}_2, \bm{Q}_3\}$, $m^z_{\bm{Q}}$, and the scalar chirality, $\chi_0$, obtained from the simulated annealing in the effective spin model on the triangular lattice for $L=96$. 
}
\end{center}
\end{figure}

\begin{figure}[t!]
\begin{center}
\includegraphics[width=0.88\hsize]{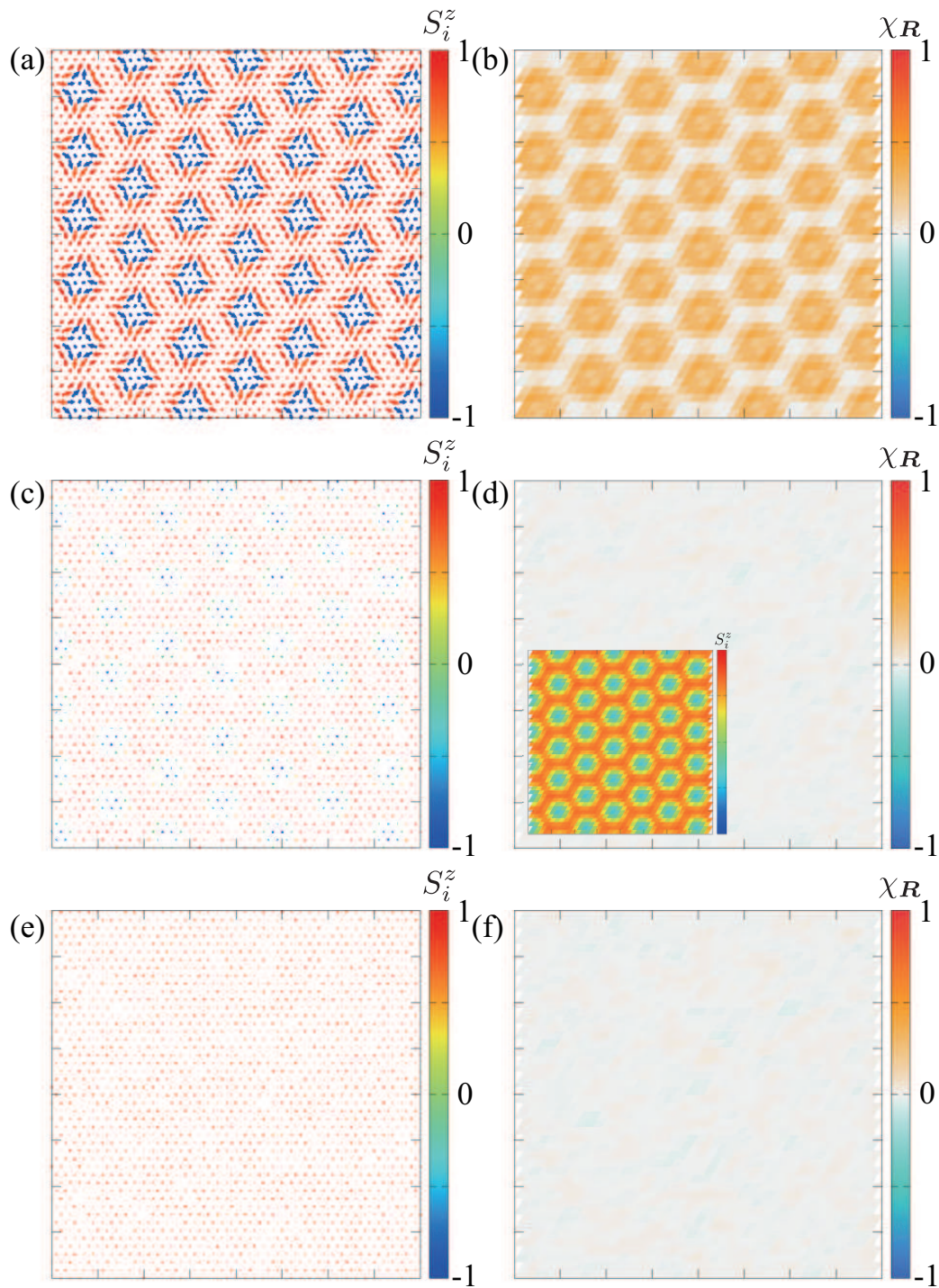} 
\caption{
\label{fig:spin_JK}
Snapshots of the real-space spin [(a), (c), and (e)] and chirality [(b), (d), and (f)] configurations in the (a), (b) SkX at $T=0.01$, (c), (d) bubble crystal at $T=0.55$, and (e), (f) paramagnetic state at $T=0.65$ obtained by the simulated annealing. 
In (a), (c), and (e), the arrows represent the magnetic moments and the color represents the $z$-spin component. 
In the inset of (d), the contour plot of the $z$-spin component in (c) is shown. 
}
\end{center}
\end{figure}

Figure~\ref{fig:Tdep_JK} shows the temperature dependence of $m^z_{\bm{Q}}=\sqrt{\frac{1}{N}\sum_{\nu} S^z(\bi{Q}_{\nu}) }$ and $\chi_0$ obtained by simulated annealing. 
In contrast to the result in figure~\ref{fig:Tdep}, we always obtain the single domain in each magnetic phase, which might be due to the interactions limited at particular $\bm{Q}_\nu$. 
The phase sequence while changing the temperature is similar to that in the Kondo lattice model in figure~\ref{fig:Tdep}. 
The SkX is stabilized at low temperatures, the bubble crystal is stabilized at intermediate temperatures, and the paramagnetic state appears at high temperatures. 
We show the real-space snapshots of the spin and chirality in each phase in figures~\ref{fig:spin_JK}(a)-\ref{fig:spin_JK}(f), which well correspond to those in the Kondo lattice model. 
In other words, the separation of the transition temperature in terms of spin and chirality degrees of freedom also arises in the effective spin model, which indicates that the effective spin model qualitatively captures the finite-temperature behavior of the Kondo lattice model. 
Thus, the model in (\ref{eq:HamJK}) is a minimal model to describe the finite-temperature transition from the SkX to the bubble crystal. 
In fact, we did not obtain such a phase transition when we omit any of $\tilde{J}$, $K$, $A$, and $H$.

\section{Summary}

To summarize, we have investigated the stability of the SkX in the Kondo lattice model with the easy-axis single-ion anisotropy on the centrosymmetric triangular lattice. 
By performing KPM-LD simulations for the large system size, we find that thermal fluctuations drive the SkX to the bubble crystal. 
Although both states are characterized by the triple-$Q$ states where the cores form the triangular lattice, only the SkX has the net scalar chirality inducing the topological Hall effect, while the bubble crystal has a collinear spin texture. 
We also show that finite-temperature transition between the SkX and the bubble crystal is found in the effective spin model derived from the Kondo lattice model, which suggests that the interplay between the biquadratic interaction and the single-ion anisotropy plays an important role in the transition.
As the similar phase transition is also found in frustrated magnets~\cite{Hayami_PhysRevB.93.184413}, the transition between the SkX and the bubble crystal is a universal feature in centrosymmetric magnets with the strong magnetic anisotropy.

\ack
This research was supported by JSPS KAKENHI Grants Numbers JP19K03752, JP19H01834, JP21H01037, and by JST PRESTO (JPMJPR20L8). 
Parts of the numerical calculations were performed in the supercomputing systems in ISSP, the University of Tokyo.

\bibliographystyle{iopartnum}
\bibliography{ref}

\end{document}